\documentclass[a4paper, 12pt]{article}
\usepackage{epsfig}

\newcommand{\be}{\begin{equation}}
\newcommand{\ee}{\end{equation}}
\newcommand{\ba}{\begin{eqnarray}}
\newcommand{\ea}{\end{eqnarray}}
\newcommand{\bi}{\begin{itemize}}
\newcommand{\ei}{\end{itemize}}

\newcommand{\lam}{\lambda}
\newcommand{\re}{\mathop{\rm Re}}

\renewcommand{\>}{\rangle}  

\newcommand{\la}{\label}

\begin{document}
\begin{titlepage}
\begin{flushright}
\end{flushright}
\begin{centering}
\vfill
 
{\bf \large GLUEBALL REGGE TRAJECTORIES \\IN (2+1) DIMENSIONAL  GAUGE THEORIES }

\vspace{1.5cm}

Harvey~B.~Meyer\footnote{meyer@thphys.ox.ac.uk},
Michael~J.~Teper\footnote{teper@thphys.ox.ac.uk}

\vspace{0.8cm}

Theoretical Physics, University of Oxford, 
1 Keble Road,\\ Oxford, OX1 3NP, United Kingdom

\vspace*{3.0cm}

\end{centering}
{\bf Abstract.-}
We compute glueball masses for even spins ranging 
from 0 to 6, in the  $D=2+1$ $SU(2)$ lattice gauge theory.
We do so over a wide range of lattice spacings, and this 
allows a well-controlled extrapolation to the continuum limit.
When the resulting spectrum is presented in the form of a 
Chew-Frautschi plot we find that we can draw 
a straight Regge trajectory going through the lightest glueballs 
of spin 0, 2, 4 and 6. The slope of this trajectory is small
and turns out to lie between  the predictions of the adjoint-string 
and flux-tube glueball models. The intercept we find, 
$\alpha_0 \sim -1$, is much lower than is needed for
this leading trajectory to play a `Pomeron-like'
role of the kind it is often believed to play in $D=3+1$. We elaborate 
the Regge theory of high energy scattering in 2 space dimensions, 
and we conclude, from the observed low intercept, 
that high-energy glueball scattering 
is not dominated by the leading Regge pole exchange,
but rather by a more 
complex singularity structure in the region 
$0\leq {\rm Re}\lam\leq \frac{1}{2}$ of the complex angular 
momentum $\lam$ plane. We show that these conclusions do not change
if we go to larger groups, $SU(N>2)$, and indeed
to $SU(\infty)$, and we contrast all this with our very 
preliminary calculations in the D=3+1 $SU(3)$ gauge theory.
\noindent 
\vfill

\vspace*{1cm}
\vfill
\end{titlepage}

\setcounter{footnote}{0}

\section{Introduction}
\label{section_intro}

On a Chew-Frautschi plot ($J$ versus $m^2$) the experimentally 
observed mesons and baryons appear to lie on (nearly) linear 
and parallel Regge trajectories, $J=\alpha_0 +
\alpha^{\prime} m^2$, with the exchange of the
corresponding Regge poles (normally) dominating any high energy
scattering that involves the exchange of non-trivial quantum 
numbers 
\cite{kaidalov}. 
The total cross-section, on the other hand, is related 
by unitarity to forward elastic scattering and this is dominated 
by the `Pomeron' which carries vacuum quantum numbers
\cite{kaidalov,book,landshoff}. 
The Pomeron
trajectory is qualitatively different from other Regge trajectories
in that it appears to be much flatter ($\alpha^{\prime}$ much smaller)
and it is not clear what physical particles correspond to integer
values of $J$. There are long-standing speculations that
these might be glueballs. Thus if one were to calculate the
mass spectrum of the SU(3) gauge theory one could investigate
whether the glueball masses fall on linear trajectories and
whether these trajectories have the properties of the Pomeron. 

Of course one does not expect the leading glueball trajectory
to be exactly like the Pomeron since in the real world 
there will be mixing between glueballs and flavour-singlet 
$q\bar{q}$ mesons; just as the usual meson/baryon 
trajectories cannot be exactly linear because the higher-$J$ 
mesons are unstable. It is only in the limit of $SU(N\to\infty)$
that one can expect exactly linear trajectories (no decays) and the 
leading glueball Regge trajectory to be the Pomeron (no mixing).
However there are good heuristic arguments to believe that 
$QCD_{N=3}$ is close to $QCD_{N=\infty}$
\cite{largeN}, 
and lattice calculations
support this idea to the extent that they have shown that for
pure gauge theories even $N=2$ is `close to' $N=\infty$
\cite{teper98}. 

Although it is now easy to calculate the lightest masses of a
pure gauge theory using lattice Monte Carlo simulations,
calculating masses of higher $J$ states is more subtle
because a cubic lattice respects only a small subgroup
of the full rotation group and each representation of this
subgroup contains states that correspond to different $J$ 
in the continuum limit. This difficulty is compounded
by the fact that the higher $J$ states, being heavier,
are more difficult to calculate accurately. In a recent paper 
\cite{hspin}
we have developed techniques for identifying
states of arbitrary $J$ on a lattice and we are now in the process
of applying these techniques to determine if glueballs fall
on linear Regge trajectories and if so whether the leading
trajectory has the characteristics of the Pomeron.

In this paper we address this question in the context of
the  $D=2+1$ $SU(2)$ gauge theory. At first glance this may seem
far removed from the case that is of immediate 
physical interest, $SU(3)$ in $D=3+1$, but the fact
that the computations are much faster in $D=2+1$ than
in $D=3+1$ means that we can expect to obtain more accurate 
results, more quickly, and so test the efficiency of our 
approach. Moreover, at closer inspection, one finds
that  $D=2+1$ non-Abelian gauge theories resemble those in 
$D=3+1$ in a number of relevant respects. They become free
at short distances, the coupling sets the dynamical
length scale, and the (dimensionless)  coupling becomes 
strong at large distances. They are linearly confining, and
the confining flux tube appears to behave like a simple
bosonic string at large distances
\cite{blmtstring}. 
The light glueball spectrum more-or-less fits the expectations 
of a flux loop model, just as it does in 3+1 dimensions
\cite{isgur,moretto,johnson}. 
Moreover, as we shall see below, the same simple models that 
serve to predict linear glueball Regge trajectories in 
$D=3+1$, predict their existence in the case of $D=2+1$.
We also note that the link
to string theories (at least at large $N$) can be made
in $D=2+1$ just as in $D=3+1$~\cite{witten}. For all these reasons
we believe that our exercise is of significant theoretical 
interest.  

At a more heuristic level, one is motivated to search
for a Pomeron trajectory where one has high energy
cross sections that are roughly constant in energy.
Although the scattering of glueballs has not been observed 
experimentally,
one's intuition is that they will behave as `black discs',
just like the usual mesons and hadrons, and so it makes
sense to speculate that the Pomeron might be the leading
(glueball) Regge trajectory in the D=3+1 SU(3) gauge theory.
We do not expect this to depend strongly on the number of
colours, so it should be a property of all SU(N) gauge theories.
Finally, since we can think of no obvious reason why going
from 3 to 2 spatial dimensions should prevent
colliding glueballs from having roughly constant cross-sections 
at high energies -- although as `black segments' rather than as
`black discs' -- we believe it makes sense to search for
something like the Pomeron in D=2+1 SU(N) gauge theories.

Although our primary aim here is to see if we can 
learn something about the Pomeron from our lattice
glueball calculations, the fact is that our results
can also be used both to test models and to test the accuracy 
of approximate analytic calculations. An example of the
former is the flux loop model for glueballs 
\cite{isgur}
which has recently been compared to the low-$J$ glueball mass
spectrum of $D=2+1$ $SU(N)$ gauge theories
\cite{johnson}.
An example of the latter is  transverse lattice light 
front quantisation
\cite{dalley}.
(Which has some problems similar to ours in determining
states of arbitrary $J$.)  Another example is 
the conjectured duality between supergravity and large $N$ gauge 
theories
\cite{maldacena}
and its generalisation to the non-supersymmetric 
case
\cite{witten}.
Extensive quantitative calculations have been undertaken
\cite{terning} 
to compute glueball masses in (2+1) and (3+1) dimensions using the 
AdS/CFT correspondence. The spectra corresponding to the quantum numbers
$2^{++},~1^{+-},~1^{--},~0^{++},~0^{--}$ and $0^{-+}$ have been computed.
It would be interesting to have a computation of the higher spin states
in order to compare them to the new lattice data and to see if they lie 
on straight Regge trajectories. That would require the inclusion of stringy 
corrections to the low-energy effective action. However, because
these higher spin states derive from string dynamics, one can expect them
to lie on straight Regge trajectories.

The contents of this paper are as follows. In the next
section we remind the reader of two simple, but plausible, 
models for glueballs in 3+1 dimensions and show how they 
directly carry over into 2+1 dimensions where they also
predict linear Regge trajectories with small slopes. 
In the following section
we discuss high energy scattering. We begin
by showing how Regge theory can be applied in two space 
dimensions (with the details relegated to Appendices A and B).
We then turn to perturbative Pomeron calculations,
where there has been a great deal of work during the last
decade, and investigate what happens  when one moves from
3 to 2 space dimensions. Here we start with the old Low
model, and the more recent dipole-dipole scattering approach,  
then we move on to the issue of gluon Reggeisation and finish
with the BFKL Pomeron. Having established the general
theoretical and phenomenological background to the problem,
we turn to our lattice calculation of how glueball masses
increase with their spin $J$. The calculation is novel in
two respects. Firstly one has to surmount the problems posed
by the limited rotational invariance of a square lattice.
This we do using the techniques recently developed in 
\cite{hspin}.
Secondly, the fact that the higher-$J$ glueball masses
are considerably heavier makes it difficult to calculate
their masses from Euclidean correlators. Here we use a
multi-level algorithm recently developed 
in
\cite{error},
which extends to glueballs earlier work 
\cite{string}
for Polyakov loops. Using these techniques we obtain quite accurate
continuum extrapolations of the $J=0,2,4,6$ glueball masses.
We find that the lightest glueballs of even $J$ lie
on a linear trajectory in a Chew-Frautschi plot of
$J$ versus $m^2$, and that the slope is small just as one
would expect for a Pomeron pole. However the intercept
is much too low to provide a constant high energy cross-section,
and we discuss the physical implications of this result.
Finally we present some results for the leading glueball 
trajectory in $SU(N>2)$ gauge theories, showing that there
is no qualitative change as $N$ varies from 2 to $\infty$.

We conclude with a summary, a discussion of how our calculations 
should be improved upon, and a mention of the preliminary 
results we have obtained in a similar study of 
the quenched 3+1 dimensional $SU(3)$ theory.

\section{Glueball models and the large $J$ limit}
\label{section_models}

In the standard valence quark picture, a high $J$ meson
will consist of a $q$ and $\bar{q}$ rotating rapidly around
their common centre of mass. For
large $J$ they will be far apart and the chromoelectric
flux between then will be localised in a flux tube which
also rotates rapidly, and so contributes to $J$. In a
generic model of such a system (see e.g. 
\cite{perkins})
a simple calculation (that we shall repeat below) shows
that the spin and mass are related linearly, 
$J=\alpha_0 + \alpha^\prime M^2$, and that the slope
is related to the tension $\sigma_f$ of the
confining string as  $\alpha^\prime = 1/2\pi\sigma_f$.
(The subscript $f$ indicates that the charges and
flux are in the fundamental representation. We will often 
follow convention and use $\sigma\equiv\sigma_f$ instead.) 
If one uses a phenomenologically sensible value for
$\sigma_f$ one obtains a value of $\alpha^\prime$
very similar to that which is experimentally observed 
for meson trajectories. This picture might well
become exact in the large-$N$ limit where the
fundamental string will not break and all the
mesons are stable. 

This picture can be generalised directly to glueballs.
We have two rotating gluons joined by a rotating flux tube 
that contains flux in the adjoint rather than fundamental 
representation. This is the first model we consider below. 
However for glueballs there is another possibility that is
equally natural: the glueballs may be composed of closed
loops of fundamental flux. This is the second model
we consider. The first model is natural in a 
valence gluon approach, while the second arises naturally
in a string theory. They are not exclusive; both may contribute
to the glueball spectrum. Indeed if there are two classes of 
glueball states, each with its own leading Regge trajectory,
then this might provide an explanation for why 
experimentally there appear to be two distinct Pomeron
trajectories, the hard `Pomeron' and the `soft' Pomeron
\cite{landshoff}.
However such speculations belong to $D=3+1$ rather than to
the $D=2+1$ gauge theories that we analyse in this paper.
What is interesting for our purposes is that both models
can be motivated as easily in $D=2+1$ as in $D=3+1$
and in both cases predict (see below) linear glueball
trajectories with some Pomeron-like properties.

\subsection{Adjoint string model}
\label{subsection_as_model}

In this model a high-$J$ glueball is imagined to be composed
of two gluons joined by a `string' in the adjoint representation,
with the whole system rotating rapidly. This is a direct extension
to glueballs of the usual model for high-$J$ $q\bar{q}$ mesons
where the $q$ and $\bar{q}$ are joined by a `string' in the 
fundamental representation. The adjoint string is of course unstable, 
once it is long enough (as it will be at high $J$), but this is also 
true of the fundamental string in QCD. What is important for
the the model to make sense is that the decay width should
be sufficiently small. (Essentially that the lifetime of the adjoint 
string should be much longer than the period of rotation.)
In $SU(N)$ gauge theories, both the
adjoint and fundamental strings become completely stable as
$N\to\infty$. So if we are close to that limit the model should
make sense. Since adjoint string breaking in $SU(N)$ occurs at
$O(1/N^2)$ while fundamental string breaking in $QCD_N$ occurs 
at $O(1/N)$, one would expect the instability to be less of
a problem in the former case. Moreover there is now considerable
evidence
\cite{blmtglue,blmtuwT,pisaQ}
from lattice calculations that the $D=3+1$ $SU(3)$ 
gauge theory is indeed `close' to $SU(\infty)$, and that this 
is also the case for the $D=2+1$ $SU(2)$ gauge theory
\cite{teper98}.

The calculation of how $J$ varies with the mass $M$ of the
glueball is exactly as for the  $q\bar{q}$ case
\cite{perkins}.
That is to say, we consider the string joining the two gluons as a 
rigid segment of length $2r_0$, rotating with  angular momentum $J$ 
(the contribution of the valence gluons being negligible at high
enough $J$). The local velocity at a point along the segment is thus
$v(r)=r/r_0$ (one maximises $J$ at given $M$ if the end-points move 
with the speed of light), so that
\ba
M&=&2\int_0^{r_0} \frac{\sigma_a dr}{\sqrt{1-v^2(r)}}=\sigma_a\pi r_0\\ 
J&=&2\int_0^{r_0} \frac{\sigma_a r v(r)dr}{\sqrt{1-v^2(r)}}=
\frac{\pi}{2}\sigma_a r_0^2,
\ea
and, eliminating $r_0$, 
\be
J=\frac{M^2}{2\pi\sigma_a}
\ee
we obtain a linear Regge trajectory of slope 
$\alpha_{AS}'=\frac{1}{2\pi\sigma_a}$ where $\sigma_a$ is the adjoint
string tension. So this model predicts that the slope of the leading
glueball trajectory is smaller than that of the leading meson
trajectory by a factor $\sigma_a/\sigma_f$.
For SU(3) we know that $\sigma_a \simeq 2.25 \sigma_f$
\cite{deldarbali},
so the leading glueball trajectory will have a slope
$\alpha_{AS}' \sim 0.88/2.25 \sim 0.39 ~\mathrm{GeV}^{-2}$ 
if we input the usual Regge slope of about 
$\alpha_{R}'=\frac{1}{2\pi\sigma_f} \simeq 0.88~\mathrm{GeV}^{-2}$.
This is only a little larger
than the actual slope of the Pomeron. Thus to this extent the model
is consistent with the idea that the Pomeron is the leading glueball
trajectory, perhaps modified by mixing with the flavour-singlet 
meson Regge trajectory. Unfortunately the model cannot predict the 
intercept $\alpha_0$ of the trajectory, because it is
valid at best at large $J$.

Since in this model the rotating glueball lies entirely within 
a plane, the calculation is identical for $D=2+1$ and  $D=3+1$, 
as indeed is the motivation for the applicability of the model.
Thus it is a plausible model for the leading Regge trajectory
in the $D=2+1$ $SU(2)$ gauge theory that we investigate in this 
paper. The only difference is that in SU(2) one expects
$\sigma_a \simeq 8\sigma_f/3$ and hence an even flatter
trajectory than in SU(3).

The above simple classical calculation can be made more rigorous 
in an effective Hamiltonian approach; see
\cite{kaidalov}
for a review. There the calculation is for $D=3+1$, 
but the main conclusion remains the same in 2+1 dimensions.

\subsection{The flux-tube model}
\label{subsection_ft_model}

An `open' string model of the kind described above, is
essentially forced upon us if we wish to describe high-$J$
mesons within the usual valence quark picture. For
glueballs, however, there is no experimental or theoretical
support for a valence gluon picture. A plausible alternative
is to see a glueball as being composed of a closed loop
of fundamental flux. Such a picture arises naturally
in a string theory approach to gauge theories. A 
simple first-quantised model of glueballs as closed
flux tubes was formulated some time ago
\cite{isgur}
and has been tested with some success 
\cite{moretto,johnson}
against the mass spectrum of D=2+1 SU(N) gauge theories as 
obtained on the lattice
\cite{teper98}.

In this model the essential component is a circular
closed string (flux tube)  of radius $\rho$. There are 
phonon-like excitations of this closed string
which move around it clockwise or anticlockwise and
contribute to both its energy and its angular momentum.
The system is (first) quantised so that we can calculate, 
from a Schr\"odinger-like wave equation, the amplitude for
finding a loop of a particular radius. The phonon
excitations are regarded as `fast' so that they contribute
to the potential energy term of the equation and
the phonon occupation number is a quantum number
labelling the wave-function.
If we are interested in the lowest mass at a given $J$, as we are 
for the leading trajectory, then we want the potential energy 
\cite{johnson}
that corresponds to the minimum number of phonons
needed to provide that $J$ 
\be 
E(\rho)=2\pi\rho\sigma_f+\frac{J-c}{\rho}
\ee
and we minimise this expression with 
respect to $\rho$ (in much the same way as in bag models)
to obtain the glueball mass $M$
\be 
M = \min_{\rho} E(\rho) = (8\pi\sigma_f J)^{\frac{1}{2}}.
\ee
This corresponds to a linear Regge trajectory
\be
J=\frac{M^2}{8\pi\sigma_f}
\ee
with a slope 
$\alpha_{FT}'=\frac{1}{8\pi\sigma_f} 
= \frac{1}{4}\alpha_{R}' \simeq 0.22~\mathrm{GeV}^{-2}$,
if we input the usual value for the Regge slope
(as in Section~\ref{subsection_as_model}). This is similar to, 
but somewhat smaller than, the observed Pomeron slope.
This analysis can be readily transformed into a 
variational calculation that minimises the Hamiltonian,
without changing the final conclusion (for large enough 
values of $J$).

The above calculation was carried out for the $D=2+1$ 
version of the model. In  $D=3+1$ there are extra 
phonons corresponding to fluctuations of the loop that
are orthogonal to the plane of the loop, and in 
addition the system can acquire angular momentum through
spinning about its axis. This alters the details of the
calculation, but not the qualitative conclusion.

Finally we remark that for $SU(N>3)$ gauge theories
the fundamental string is no longer the only one that
is absolutely stable, and closed loops of these 
higher representation strings provide an equally
good model for glueballs
\cite{johnson,blmtstring}.
These extra glueballs
will however be heavier and, to the extent that
we are only interested in the leading Regge trajectory,
will not be relevant for the discussion of this paper.

\section{High-energy scattering in (2+1) dimensions}
\label{section_scatt}

We begin by describing how the familiar Regge theory
of 3 spatial dimensions can be taken over to 2 space dimensions.
We confine ourselves to a brief summary here, leaving 
details to the Appendix.

The Pomeron has been the focus of a large number of field theoretic
studies, many motivated by the `hard Pomeron' and low-$x$ physics
in deep inelastic scattering. Just as we asked how 
glueball models change when we go from 3+1 to 2+1
dimensions, it is interesting to know the predictions of these
perturbative approaches when we change dimension.
That will be the focus of the remainder of this Section.

\subsection{Regge theory predictions}
\label{subsection_regge}

The optical theorem relates the total cross section to the
dimensionless scattering amplitude $T(s,t)$ by
\be
\sigma_{tot}=\frac{1}{\sqrt{s}}~\mathrm{Im}~
T(s,t=0).
\ee 
(In two space dimensions the `cross section' has dimensions of length.)
As is shown in appendix A, it is given by the following contributions:
\be
T(s,t)=a_0(s)~~+~~{\rm background~integral}~~+~~\sum ~\left[{\rm
Regge~pole~contributions}~\propto~\left(s^{\alpha(t)}\right)\right],
\ee
where $\alpha(t)$ describes the Regge trajectory in the Chew-Frautschi plot.
This equality is based on the analytic continuation  of the partial
waves in $\lam$, the angular momentum, and on crossing symmetry. 
There are two differences with 
respect to the 3+1 dimensional case: the background integral gives a constant
contribution to the amplitude, rather than decreasing as $\frac{1}{\sqrt{s}}$;
and the s-wave exchange is not included in the Sommerfeld-Watson transform. 
In potential scattering, and even more general situations, 
it can be shown to be a branch point in the complex
$\lam$ plane at threshold (see
\cite{chadan98} 
and appendix B). 

\subsection{QCD$_2$ at high energies}
\label{subsection_QCDhighE}

We first give the simplest estimates of the color-singlet exchange
for high energy scattering. We then comment on the failure of gluon 
reggeisation and review the results of Li and Tan
\cite{tan}
for color-singlet exchange  obtained in the leading logarithm approximation.
In order to develop some intuition for 2+1 dimensional physics,
we finish with a discussion of the momentum dependence of hadronic 
structure functions.

\subsubsection{Color singlet exchange in leading order}
\label{subsubsection_singlet}

If we compute the color-singlet part of a two-gluon exchange diagram
between two `quarks' in 2+1 dimensions
(the first perturbative Pomeron model due to F. Low in 1975
\cite{low}), 
we find
\be 
A_1^{(1)}=i\alpha_s^2s  \frac{N_c^2-1}{N_c^2}\int\frac{dk}{k^2(k-q)^2},
\ee
 implying, by use of the optical theorem in 2+1 dimensions (appendix A), 
\be 
\sigma_{\mathrm{tot}}(qq\rightarrow qq)=\alpha_s^2 \frac{N_c^2-1}{N_c^2}
 \int \frac{dk}{k^4} \la{low}
\ee
The result is completely similar to the 3+1 case, except that the IR 
divergence is worse by one power -- $\sigma_{tot}$ has units of length.
The Pomeron exchange amplitude is finite
 once impact factors are introduced for the hadrons -- 
giving the incoming quarks a slight offshellness and effectively introducing
a cutoff in the integral on the right-hand side of~(\ref{low}). 

Another approach  consists in computing dipole-dipole scattering 
(for an introduction, see
\cite{dipolecture} 
and references therein). Proceding as in the 3+1 dimensional 
case, the leading order (large $N$) dipole-dipole cross-section reads
\be
\sigma_{dd}=4\alpha_s^2\int_{-\infty}^\infty \frac{d\ell}{\ell^4}
(1-\cos{\ell x_{01}})(1-\cos{\ell x'_{01}})=\pi \alpha^2 x_<^3
(3\frac{x_>}{x_<}-1)\qquad(2+1)
\ee
as compared to
\cite{dipolecture}
\be
\sigma_{dd}=2\pi\alpha_s^2 x_<^2\left(1+\log{\frac{x_>}{x_<}}\right)\qquad(3+1)
\ee
where $x_>$ ($x_<$) is the greater (lesser) of the two dipole sizes $x_{01}$
and $x_{01}'$. In both cases, we find a constant cross-section. 

To go beyond the leading contribution, several calculational schemes
 are available.
In particular, the BFKL Pomeron is obtained by keeping, order by order
 in $g^2$, only the leading logarithmic contribution in the perturbative
 expansion. The first step in calculating the amplitude for Pomeron exchange
is to establish gluon reggeisation.

\subsubsection{The issue of gluon reggeisation}
\label{subsubsection_gluon}

In the Regge limit $s\gg t \gg g^4$, where $s,t$ are the Mandelstam variables,
the usual expansion in $g^2\log{s}$ yields the 0, 1 and 2-loop amplitudes
for color octet exchange:
\ba
A_0^{(8)}&=&8\pi\alpha_s G_0^{(8)} \frac{s}{t} \\
A_1^{(8)}&=&A_0^{(8)}  \epsilon_G(t) \log{\frac{s}{k^2}}\\
A_2^{(8)}&=&\frac{1}{2}A_0^{(8)}\left(\epsilon_G(t) 
\log{\frac{s}{k^2}}\right)^2 
\ea
where $k^2 = O(t)$ and
\be
\alpha_s=\frac{g^2}{4\pi},\qquad G_0^{(8)}=\tau^a_{ij}\tau^a_{kl},
\qquad \epsilon_G(t)=N_c\alpha_s\int^{+\infty}_{-\infty}
\frac{dk}{2\pi} \frac{t}{k^2(k-q)^2}\le 0\qquad (t=-q^2)
\ee
Thus, at least formally, the gluon reggeises
\cite{BFKL2+1}:
\be
A^{(8)}=A_0^{(8)}~\left(\frac{s}{k^2}\right)^{\epsilon_G(t)}
\ee
The infrared divergence in the quantity $\epsilon_G(t)$ is linear
 (as opposed to logarithmic in 3+1 dimensions), 
and it must be so since $\alpha_s$ carries dimension of mass. 
The IR ``gluon mass'' cutoff $M$ has to be introduced, in which case
$\epsilon_G(t)=\frac{N_c\alpha_s}{M}$. Physically $M$ can be interpreted 
as a non-perturbative mass that the gluon acquires at the confining scale; 
therefore we expect $g^2/M=\mathcal{O}(1)$. This however shows that, due to
the infrared divergence, the result of the perturbative calculation has 
a linear sensitivity  to physics at the confinement scale $g^2$, 
where the perturbative expansion breaks down. 

In the Verlinde approach
\cite{verlinde} 
to high-energy scattering adopted by Li and Tan
\cite{tan}, 
gluon reggeisation fails. However, as the authors remark, 
this is not necessarily in contradiction with conventional perturbative
calculations, since the really physical quantity is the 
color-singlet exchange.

\subsubsection{The 2+1 perturbative Pomeron}
\label{subsubsection_pertPomeron}

The BFKL equation was solved exactly in the presence of a gluon mass in
\cite{BFKL2+1}. 
However when this mass is taken to zero, 
the IR divergence shows up in the fact that the BFKL exponent $\omega_0$ 
runs as $\sim\alpha_s/M$; this fact could be guessed on dimensional grounds.
Within perturbation theory, such a mass $M$ can only appear as an IR 
regulator. The situation is in radical contrast to  the 3+1 dimensional case, 
where the cancellation of  IR divergences in the BFKL equation makes
 it self-consistent.
In the detailed calculation, the simple structure of the infinite series
is spoiled in the $M\rightarrow 0$ limit 
by the re-emergence of a power dependence on $s$ at each 
order due to the IR divergences. Thus, in this framework, 
 a power-like dependence of the cross-section on $s$ in the limit of
 zero gluon mass is not possible in 2+1 dimensions.

A thorough investigation of $QCD_2$ high energy scattering 
 was undertaken by Li and Tan~\cite{tan}. In their first 
paper, they used the Verlinde approach~\cite{verlinde} to obtain a 
one-dimensional action, where they are able to compute the (finite) 
color-singlet exchange exactly. They predict a 
\be
\sigma\propto 1/\log{s}
\ee
dependence
of the total cross section on the center-of-mass energy. In a second 
publication, they re-derive this result using the dipole picture 
(\cite{dipolecture} and references therein) of 
high-energy scattering. In this case all quantities are naturally IR-safe.


\subsubsection{Deep inelastic scattering in 2+1 dimensions}
\label{subsubsection_deepinel}

A standard prediction of the BFKL Pomeron in 3+1 dimensions 
is the strong rise of the deep inelastic structure functions
at low $x$ when $Q^2$ is large (for an introduction, see
\cite{book}): 
\be
F(x,Q^2)\sim \frac{x^{-\omega_0}}{\sqrt{\log{1/x}}}
\ee
where $\omega_0=4\frac{N_c}{\pi}\alpha_s \log{2}$ is the BFKL exponent.

On the other hand,
the DGLAP equation for the evolution of the moments $M(n,Q^2)$ of the parton 
distributions leads to the behaviour
\be 
M(n,Q^2)=C_n~\left(\log{\frac{Q^2}{\Lambda^2}}\right)^{-A_n}
\ee
where the pure number $A_n$ is an ``anomalous dimension'' computed in 
perturbative QCD. The $Q^2$ dependence comes from the running of the coupling 
$\alpha_s$; in 2+1 dimensions, the equation is therefore replaced by
\be
\frac{\partial M(n,Q^2)}{\partial \log{Q^2}}=A_n \frac{\alpha_s}{Q} M(n,Q^2)
\ee
yielding the following large $Q^2$ behaviour:
\be
M(n,Q^2)=M(n,\infty)~\exp{\left(-\frac{2A_n\alpha_s}{Q} \right)}\simeq 
M(n,\infty)-2A_n\frac{\alpha_s}{Q}.
\ee
That is to say, the structure functions tend to finite constants at large $Q^2$.
The physical reason for this is that at large energy scales, the theory 
becomes free very rapidly (the effective coupling scales as  $1/E$), 
which does not allow for an evolution of the structure functions. Thus above
the confinement scale, we qualitatively expect a rapid evolution in $Q^2$ 
of the structure function toward its asymptotic value. 
Once a high $Q^2$ has been reached, 
the virtual photon $\gamma^*$ does not ``see'' more partons when
its resolution is increased, because the amplitude that they be emitted is 
suppressed by $\alpha_s/Q$.

\section{The D=2+1 SU(2) glueball spectrum}
\label{section_glueballs}

In the first part of this Section we discuss the technical
aspects of the lattice calculation. 
In the second part we present the results of our calculations.

\subsection{Technical aspects}
\label{subsection_technical}

The identification of the continuum spin of a lattice state 
requires novel techniques, as does the accurate calculation
of the mass of a heavy, high-spin state, and we discuss
these below. First however we briefly summarise the more standard
aspects of the calculation.

\subsubsection{Masses from lattice simulations}
\label{subsubsection_latticemass}

To obtain masses we calculate Euclidean correlation functions
of some well-chosen operators $\phi_i$
\be
C_{ii}(t)
\equiv
\langle \phi^{\dagger}_i(t)\phi_i(0) \rangle
=
\langle \phi^{\dagger}_i e^{-Ht} \phi_i \rangle
=
\sum_n |a_n|^2 e^{-E_n t}.
\ee
Here the $E_n$ are the energies and 
$a_n=\langle n | \phi_i | vac \rangle $. If the vacuum has trivial
quantum numbers, then only states with the quantum numbers of $\phi_i$ 
will have $a_n \not= 0$.  Suppose we have $i=1,...,n_0$ operators
of the same quantum numbers and we also calculate the 
off-diagonal correlators $C_{ij}(t)$. Then an effective way 
to calculate the lightest few states with these quantum numbers
is to perform a variational calculation using the basis
$\{\phi_i; i=1,...,n_0\}$. For a recent exposition of this standard
method see
\cite{teper98}.

To calculate such Euclidean correlation functions we use
lattice Monte Carlo methods. 
As usual one needs to ensure that the operators are smooth
and extended, so that they have a good projection onto
the lighter physical states, and we use an iterative
smearing technique for that purpose
\cite{smear}. 
We use an 
increasing number of smearing iterations (and also increase
the smearing weights) as we approach the continuum. 
All our calculations use the standard Wilson action. 
The update is a 1:3 mixture of heat-bath
\cite{hb,kenpen} 
and over-relaxation
\cite{adler} 
sweeps. Because we calculate the values 
of many operators, the `measurement' is time-consuming and
we can do a significant number of these compound sweeps between
measurements without significantly increasing the total cost of
the calculation. 
We typically perform  ${\cal O}(10^5)$ sweeps and collect the data 
in $\sim 100$ bins. Errors are estimated with a standard jackknife 
analysis.

\subsubsection{Continuum spin on a cubic lattice}
\label{subsubsection_spin}

Consider the eigenstates of the transfer matrix of the lattice 
field theory. These will belong to the irreducible representations 
of the cubic rotation group and will not, in general, possess the
rotational properties that characterise a continuum state of 
a definite spin. However as $a\to 0$ each of these states will
tend to an energy eigenstate of the continuum theory that
possesses some definite spin $J$. By continuity for sufficiently
small $a$ the rotational properties of this lattice state
will be arbitrarily close to those of a continuum state of spin $J$.
We will therefore refer to such a state not only by its lattice
representation but also by the appropriate continuum spin $J$.
To be able to do this we need to identify, for each such lattice
state, what continuum $J$ it tends to as  $a\to 0$. 
Once we know this then we can perform a standard continuum 
extrapolation of the calculated lattice masses so as to obtain 
the continuum mass of the spin $J$ glueball.

A detailed investigation into how to do this was presented in
\cite{hspin}. 
Here we shall apply a systematic version of the method that was 
presented there under the name `Strategy II'. We briefly remind 
the reader of this method.

The operators we use lie in definite lattice irreducible 
representations (IRs), and we use the variational method
\cite{var}
to extract estimates for the eigenstates (in our operator basis)
and their masses. In this way we calculate the mass of the lightest
state and of several excited states in the given lattice IR -- 
typically the number is one third of the number of operators we 
are using. To identify which $J$ each of these states tends to,
we do a simple Fourier analysis of the wave function of the 
corresponding diagonalised operator. To do this we 
measure the correlations between it and a `probe' operator that 
we are able to rotate to a good approximation by angles smaller 
than $\frac{\pi}{2}$. This provides a measurement of the wave function.
We check the rotational properties of the probe by measuring the vacuum
wave function; typically it is found to be isotropic at the one percent level,
which is taken as evidence that the probe has good rotational properties.
We are mainly interested in the dominant component and 
the fact that there is an uncontrolled uncertainty of a few percent in the 
measurement of the wave function does not impede significantly the procedure
of determining the spin of a state in the continuum limit. This is so because
one is usually discriminating between a constant and a change of sign 
behaviour of the wave function. In principle, 
the continuum extrapolation of these coefficients should determine uniquely
the spin of the glueball. In practice, it usually turns out that there is 
already a very  dominant coefficient at finite lattice spacing -- 
except when a crossing of states occurs -- as we shall see in an 
example later on.

\subsubsection{The multi-level algorithm}
\label{subsubsection_multilevel}

A state with high spin will generically have a large mass, and the value of
the corresponding correlation function can be very small. 
Indeed, even if we have a perfect operator the value of the correlator
will be $O(e^{-aM_J n_t})$ at $t=an_t$ and if $aM_J\gg 1$
then it will be in danger of being swamped by the statistical 
errors. In practice it is not possible to restrict oneself to
values of $a$ such that $aM_J\ll 1$ for all values of $J$ that
are of interest. Especially so because we need a range of $a$ that
is large enough for a well-controlled continuum extrapolation.

To deal with this problem we make use of a recently proposed error 
reduction algorithm for glueball correlators
\cite{error}. 
It proved very useful on all but the smallest lattice spacings 
(i.e. on $\beta=6,~7.2,~9,~12$). For $6\le \beta \le 9$, we 
used $\mathcal{O}(500)$ sub-sweeps, while we decreased their number to 50 
at $\beta=12$. These sub-sweeps are done on sub-lattices which represent 
``time-blocks'' of width 4. Our experience is that it is more efficient
to do all these sweeps at one fixed time-block (that is, a 2-level algorithm)
rather than splitting up the sweeps between width 2 and width 4 blocks
(that is, a 3-level algorithm). This was noted previously in
\cite{string}, 
where a multi-level algorithm was applied to Polyakov loop measurements,
and suggested in
\cite{email}.

The choice of the number of sweeps was done on the basis of the thumb-rule 
$n_{sweeps}=e^{m\Delta t}$, where in our case $\Delta t=4a$ and the 
algorithm was optimised for the spin 4 glueball, that is $m=m_{4}$.
Indeed when one is measuring large portions of the spectrum, a compromise
 has to be made. For the lighter states, it is more efficient to do few
sub-sweeps, while the heavier states require more of them. That is, for 
the same computer time, we could have measured the lightest glueball mass 
more 
accurately had we not used a 2-level algorithm, but we would then have far
less accurate and reliable results for the spin 4 and 6 glueballs -- the main
goal of the present simulations. 

The multi-level algorithm enables us
 to apply the variational method on the correlation 
matrices at 2 and 3 lattice spacings even on the coarsest lattice. 
Very often, when the correlators are measured in the traditional way, 
the noise that dominates the signal of the heavier states spoils 
the positivity of the correlation matrices 
if the method is not applied between 0 and 1 lattice spacing,
thus impeding the variational calculation.
In our case however the very massive eigenstates benefit most from the 
sub-averaging procedure and the positivity is maintained. Usually the
orthogonalised operators have reached their mass plateaux (within error bars)
 by three lattice spacings for the data we present in this paper; this 
increases our confidence in the reliability of the variational method.

At $\beta=18$, the correlation
length $1/\sigma\simeq 12a$; that would be the natural width in the Euclidean
time direction over which to do sub-sweeps. However that would mean 
extracting the masses at time separations of that order. It turns out that
our smeared operators have  sufficiently good overlaps onto the physical 
states to reach a plateau far earlier than 12 lattice spacings. For that 
reason, we did not apply the multi-level algorithm to that case. Nevertheless
we regard this as a consequence of the improved smoothness 
of the operators close to the continuum rather than as a
defect of the multi-level algorithm.

\subsection{Results}
\label{subsection_results}

In Table~\ref{massdata} we list the values of the masses we
calculate on $L^3$ lattices at various values of $\beta=4/ag^2$.
The masses are in lattice units and are labelled both by
the lattice IR to which they belong, and by the spin $J$
of the state to which they tend in the continuum limit.
The latter assignment is achieved as described above,
and an explicit example will be given below.
We also have calculated the confining string tensions as indicated.
The string tension provides a natural dynamical length scale
$\xi \equiv 1/\surd\sigma$ which tells us how small the lattice
spacing $a$ is, $a/\xi = a\surd\sigma$, and how large our
lattice size, $aL$, is, $aL/\xi = aL\surd\sigma$.

Since high-$J$ states are expected to be very extended, it is
important to check that our $J=0,2,4,6$ mass estimates are
not subject to large volume corrections. This is the first
thing we do in this subsection. We then give an explicit example
of the Fourier coefficients which we use to identify the $J$
of the state. Finally we discuss the extrapolation to the
continuum limit.

\subsubsection{Finite volume effects}
\label{subsubsection_volume}

As one can see from Table~\ref{massdata}, the spatial size that we
use for most of our calculations satisfies $aL\surd\sigma \sim 4$. 
This choice was based on earlier finite volume studies
\cite{teper98}
that showed that it appeared to be  large enough for 
the lightest glueball states.
In particular, on such a volume the lightest state of two periodic flux 
loops (which can couple to local glueball operators) will be heavier 
than the lightest few  $A_1$ states and the lightest $A_3$ state. 
In this paper, however, we are interested in higher spin states that 
may be significantly more extended than these lightest states, 
so it is important to check for finite volume corrections by performing 
at some $\beta$ the same calculations on much larger volumes.
We do this at $\beta=7.2$, where the spatial extent of our comparison 
volume is twice as large. In addition we perform a more limited
comparison at a finer lattice spacing, $\beta=12$, on a comparison
lattice that is about $30\%$ larger.

We see from Table~\ref{massdata} that there is in fact no significant 
change in any of the masses listed when we double the lattice size
from  $aL\surd\sigma \sim 4$ to $aL\surd\sigma \sim 8$
at $\beta = 7.2$. In particular this is true for the $J=4$ and $J=6$
states where our concern is greatest. We note also that on the
$L=40$ lattice a state composed of two periodic flux loops will
have a mass $am_T \sim 2 La^2\sigma \simeq 3.45$ which is much
heavier than any of the masses listed and so it will not be a
source of finite volume corrections there. From the comparison
we infer that these `torelon' states cause no problem on
the $L=20$ lattice even though their mass $am_T \sim 1.7$ is light
enough for it to mix with the states of interest. This tells us
that in general such mixing will not be important even where
it is possible. This is consistent with the observation
\cite{teper98} 
that in many respects $SU(2)$ is close to $SU(\infty)$,
since in the latter case the mixing will vanish.

The more limited finite volume check at $\beta=12$ also
shows no significant volume variations for the $J=4$ and $J=6$
states, as well as the $J=0$ states. On the other hand
there are some anomalies in the $J=2$ part of the $A3$ spectrum.
Since on these two lattices the `torelon' state has a mass
$am_T \sim 0.9$ and $am_T \sim 1.2$ respectively it is possible
that it is mixing with some nearly degenerate $A3$ glueball 
states on both the lattice sizes, and that this explains the
anomalies. However we also remark that these calculations were 
performed at an early stage (unlike those at $\beta = 7.2$)
and a different basis of operators was used on the two
volumes. Thus it is not clear how seriously we should take
the discrepancies observed in the $A3$ sector.

\subsubsection{The Fourier coefficients}
\label{subsubsection_fourier}

As described in Section~\ref{subsubsection_spin}, 
we determine the `$J$' of a lattice
energy eigenstate by a Fourier decomposition of its wavefunction.
For example for a state in the trivial $A_1$ lattice IR we have
\begin{equation}
|\psi_1\>=\sum_{n\geq 0} c_n |(j=4n)^+\>|_{lat} \quad {\rm with } \quad 
\sum_n |c_n|^2=1 .
\label{se}
\end{equation}
The Fourier decomposition is performed using as a probe a 
loop $\cal{O}$ for which we have a number of other loops 
${\cal{O}}_\theta$ that are (to a good approximation) copies 
of  $\cal{O}$ rotated by angles $\theta$ that 
are not multiples of $\pi/2$:
 \begin{equation}
\langle vac |{\cal{O}}_\theta |\psi_1\> 
= 
\sum_{n\geq 0} c_{\cal{O}} c_n e^{i 4n\theta}
\end{equation}
(where $c_{\cal{O}}$ is independent of $n$ in the limit
where ${\cal{O}}_\theta$ is an exact rotated copy of $\cal{O}$).
Note that since the angular resolution is $O(a)$ one can 
attempt to resolve spins up to $J_{max} \sim O(1/a)$ in this way. 
In Table~\ref{coefdata} we show the Fourier coefficients calculated
at the lattice spacings $\beta=7.2,~9,~12,~18$.
The table shows the normalised $|c|^2$ coefficients corresponding to the 
spin that the state is assigned in the continuum limit. We see that the 
states that become $0^+$ have very isotropic wavefunctions even 
at the finite lattice spacings considered. The spin 4 coefficients of the 
spin-4-to-be states vary a lot more. Let us look at the fundamental spin 4 
glueball in more detail.

The coefficient is very close to one at $\beta=7.2,~9$ and $18$, but shows
a big dip at $\beta=12$. We attribute this to the crossing of the lightest
spin 4 state and the $0^{+***}$. Indeed looking at the masses in 
table~\ref{massdata}, we observe that these two states are always 
nearly degenerate, the spin $4$ being slightly heavier on the coarse lattices
and slightly lighter on the finer lattices, while they are closest precisely
at $\beta=12$. As was pointed out in
\cite{hspin}, 
an ``accidental'' 
degeneracy like this automatically  leads to maximal quantum mechanical 
mixing between the states, since there is no lattice symmetry to
prevent that. Taking this into account, the observed evolution of the
Fourier coefficient is not implausible.

At  $\beta=6$ we did not perform a
Fourier decomposition but rather chose the value of $J$ using
the level ordering already established for the other values of  $\beta$.
The reason we did not perform such a decomposition is that this
calculation was originally intended as a test of the multi-level
algorithm rather than as a contribution to the present study.
However given its accuracy it seemed wasteful not to use it.
We have a less accurate older study 
\cite{hspin}
where the Fourier decomposition was performed and that supports our
spin identification. In addition the mass of the identified
$4^{-}$ state, and the fact that it should be (nearly)
degenerate with the $4^+$ leaves no doubt about the correctness 
of the $J=4$ identification in this case as well.

\subsubsection{Continuum extrapolation}
\label{subsubsection_continuum}

Our continuum extrapolation of the states in the $A_1$ and $A_3$ 
representations is entirely conventional and follows 
\cite{teper98,hspin}. 
We plot the glueball masses in units  of the string tension as a function
of $\sigma a^2$, and attempt a linear fit
\begin{equation}
{{am_G(a)}\over{a\surd\sigma(a)}}
\equiv
{{m_G(a)}\over{\surd\sigma(a)}}
= 
{{m_G(0)}\over{\surd\sigma(0)}} + c a^2 \sigma.
\end{equation}
This is motivated by the fact that we know the leading lattice 
correction to be $O(a^2)$ for the plaquette action. 
If such a fit has a bad confidence level, then we remove the coarser 
lattice data until a good fit is obtained. We require
that at least three points are left.

Fig.~\ref{extrapolfig} shows the actual extrapolation for the lightest
state of each spin. We observe as in
\cite{hspin} 
that the evolution is weak.
Note also that thanks to the error reduction technique employed, the 
error bars corresponding to the coarser lattices do not appear very
much larger than those associated with finer lattices.
Table~\ref{contdata} gives the continuum spectrum in units of the 
string tension, as well as the confidence level and the number of
different lattice spacings included in the fit. For the fundamental
states of spin 0, 2, 4 and 6, the confidence levels are good and 
include all five lattice spacings. 

Not surprisingly, the second and third excited states 
have less reliable extrapolations. 
A slight dependence of the masses obtained for the excited states
 on the basis used in the variational method is likely to be the cause of 
the stronger spread of the lattice data in the extrapolation plot. Notice
for instance that all four states shown on Fig.~\ref{extrapolfig} appear 
slightly lighter at $\beta=18$ than at the other lattice spacings.

\section{SU($\mathrm{\bf N>2}$)}
\label{section_allN}

As we remarked in Section~\ref{section_intro}, it is only in the
$N\to\infty$ limit, where all glueballs become stable, that
one can hope to identify the ideal linear Regge trajectory.
In principle all one needs to do is to repeat the
above SU(2) calculation for $N=3,~4,~5,..$. We know from
\cite{teper98}
that the approach to $N=\infty$ is rapid so that the first
few values of $N$ should suffice for a good extrapolation
to all values of $N$. However while such a calculation is
certainly feasible, it is beyond the very limited computational 
resources currently available to us. Fortunately we
are able to finesse this practical problem using the fact that
it has been shown in
\cite{hspin}
that the lightest state in the lattice $A_2$ IR, which contains
$J^P=0^{-+},~4^{-+},~8^{-+}$, is in fact the 
$4^{-+}$ rather than the $0^{-+}$. This confirmed earlier
suggestions, based on an analysis of the predictions of
flux tube models
\cite{johnson},
that the lightest $0^{-+}$ should be much more massive
than the lightest observed $A_2$ state, while the latter was
consistent with the model prediction for the lightest
$4^{-+}$ state. Due to parity doubling in $D=2+1$
this mass is the same as that of the $4^{++}$
(in the infinite volume continuum limit). Thus we 
can use the lightest states in the $A_1$, $A_3$ and $A_2$
lattice representations, as calculated for various
$SU(N)$ groups in 
\cite{teper98},
to provide us with the lightest $J=0,~2,~4$ glueball masses.
(Note that this means that the masses labelled in the tables of
\cite{teper98}
as being those of the lightest  $0^{-+}$ should in fact be
relabelled as being the lightest  $4^{-+}$.)
This is more limited than our explicit SU(2) calculation
where we also identify the $J=6$ glueball, but it is
adequate given the presumption that there will be no
qualitative change as we increase $N$ from $N=2$.

The assumption that  for $SU(N>2)$ the lightest  $A_2$ state is 
the $4^{-+}$ is very reasonable given that this is so in SU(2)  
\cite{hspin}
and that it is predicted to be so by generic flux loop models
\cite{johnson}.
Nonetheless it is an assumption and should be checked.
We have therefore performed such a check in the $SU(5)$ case, 
at $\beta=64$, $L=24$, where $\sigma^{-1/2}\simeq 6a$. 
Using a 16-fold rotated triangular probe operator reveals
that the wave function of our best $A_2$ operator, measured at 
a Euclidean time separation of one lattice spacing,
behaves like $\sin{4x}$, with a normalised 
coefficient consistent with 1 at the few percent level.
This confirms the correctness of our assumption.

\section{Physical discussion}
\label{section_physics}

We begin by asking what our glueball spectrum tells us about
the nature of the leading glueball Regge trajectory, both 
for $SU(2)$ and for larger $N$. We then compare what we find
to the predictions of the simple glueball models in 
Section~\ref{section_models}.
Finally we discuss what role this trajectory will play
in high energy scattering. 

\subsection{The glueball spectrum in a Chew-Frautschi plot}
\label{subsection_trajectory}

In Fig.~\ref{chewfig} we plot our continuum $SU(2)$ glueball spectrum 
in a  Chew-Frautschi plot of ${m^2}/{\sigma}$ against the spin $J$.
We see that the lightest $J=0,~2,~4,~6$ masses appear to lie on
a straight line. If we fit them with a linear function
$J=\alpha(t)$, where $\alpha(t)=\alpha_0+\alpha't$ and $t=m^2$,
then we obtain  
\be
2\pi\sigma\alpha_{(m)}'=0.322(16)\qquad \alpha^{(m)}_0=-1.18(11)
\ee
with a confidence level of $65\%$. (If we drop the $J=0$ state
from the fit, the errors become somewhat larger, but the
trajectory is essentially the same.) Thus we reach the remarkable 
conclusion that the lightest glueballs of spin $J$ fall 
on a linear Regge trajectory. This is the leading trajectory, 
hence the index $(m)$ standing for `mother trajectory'.

Although there is  more uncertainty in establishing the excited 
spectrum, particularly in the spin 2 sector where finite volume 
effects are not completely understood, we also fit the $0^{+*}$, 
$2^{**}$ and $4^*$ states to a straight line and find
\be
2\pi\sigma\alpha_{(d)}'=0.265(36)\qquad \alpha^{(d)}_0=-2.20(44)
\ee
with a confidence level of $93\%$. This `daughter' trajectory
is approximately parallel to the leading one, and its
intercept is down by about one unit.

As we explained in Section~\ref{section_allN}, we can also say 
something about the leading Regge trajectory in $SU(N>2)$ gauge 
theories, if we use the masses calculated in
\cite{teper98}
and relabel as $4^{-+}$ the state that is labelled there as the
lightest $0^{-+}$. Since the $4^{-+}$ and $4^{++}$
are degenerate in the (infinite volume) continuum limit,
this gives us the  $0^{++}$, $2^{++}$ and $4^{++}$
continuum masses for $N=2,~3,~4,~5$. A linear fit,
$J=\alpha(m^2)=\alpha_0+\alpha'm^2$, works in all cases and yields:
\begin{center}
\begin{tabular}{|c|c|c|c|}
\hline
\cite{teper98} data & $2\pi\sigma\alpha_{(m)}'$ & $\alpha^{(m)}_0$ & conf. lev.          \\
\hline
$N=2$ &  0.324(15) & -1.150(75) & $89\%$ \\
$N=3$ &  0.384(16) & -1.144(71) & $54\%$ \\
$N=4$ &  0.374(18) & -1.068(75) & $71\%$ \\
$N=5$ &  0.372(22) & -1.036(88) & $86\%$ \\
\hline 
\end{tabular}
\end{center}
We have also included the result for $SU(2)$ and we note that the parameters 
of the trajectory are in very good agreement with the data of this paper.
It is clear that for all the number of colors available, the linear fit 
has a very good confidence level.

We conclude that all $SU(N)$ gauge theories possess approximately linear
Regge trajectories, with slopes 0.3---0.4 in units of $2\pi\sigma$,
and intercepts close to -1, which appear to be approaching that
value as $N\to\infty$.

\subsection{Comparison to glueball models}
\label{subsection_datamodel}

As we saw in Section~\ref{section_models}, the closed flux 
tube model of glueballs
predicts a leading Regge trajectory that is linear, with a slope
that is independent of $N$:
\be
2\pi\sigma~\alpha_{FT}'=\frac{1}{4}\qquad \forall N
\ee
The adjoint string model also predicts a linear Regge trajectory
but with a slope  $2\pi\sigma_a~\alpha_{AS}'=1$ that in general 
depends on $N$ through the $N$ dependence of $\sigma_a/\sigma$. 
Lattice calculations 
\cite{deldarbali}
(and some theoretical arguments
\cite{simonov})
support a dependence that is close to Casimir scaling, 
\be
{{\sigma_a}\over{\sigma}}
=
\frac{C_A}{C_F}
=
2\frac{N^2}{N^2-1}.
\ee
Assuming this, the slope predicted by the adjoint string model becomes
\ba
N=2:& &\qquad2\pi\sigma~\alpha_{AS}'=\frac{3}{8} \\
N=\infty:& &\qquad 2\pi\sigma~\alpha_{AS}'= 2\times(2\pi\sigma~\alpha_{FT}')
=\frac{1}{2}
\ea

Interestingly, for all the $N$ considered, the lattice result 
for $\alpha'$ is almost exactly at the midpoint between the
two model predictions. We illustrate this fact in Fig.~\ref{a1}. 
We might speculate that even if both models are valid, 
thus producing two glueball trajectories with different slopes,
at finite $N$ mixing will deform these trajectories
from exact linearity and that such a deformation will be
greatest at some lower $J$ where the states of the two
trajectories are closest and also where we perform
our calculations. 

We observe that the intercept of the leading Regge trajectory that
we have obtained is close to -1, and becomes perhaps even closer 
at larger $N$, as we see if we plot
\be
\Delta\equiv-(\alpha_0+1),
\ee
in Fig.~\ref{a0}. Now we recall that in D=3+1
\cite{book} 
the fact that the color-singlet amplitude is, order by order, 
down by a factor of $N\alpha_s$ with respect to the color octet 
exchange amplitude (see Section~\ref{subsection_QCDhighE}) 
can be interpreted as coming from an expansion of the signature
suppression factor around $J=1$, using the fact that in this picture
$\alpha(t=0) = 1 +O(\alpha_s)$. A similar argument will work in
D=2+1 as long as $\alpha(t=0)$ is near an odd integer. We assume
that $\alpha(t=0) \simeq 1$ is disfavoured since it would
lead to a rapidly rising cross-section, and so the next
possibility would be  $\alpha(t=0) \simeq -1$, as observed in
our calculations. This line of reasoning relies on the idea that 
the perturbative calculation remains qualitatively valid even as 
$t\rightarrow0$, which is of course not guaranteed.

At finite $N$ Regge trajectories are not expected to rise linearly at 
arbitrarily large $t=m^2$. In particular
we should  expect that due to mixing between high spin glueballs
and multi-glueball scattering states, for which
\be
\alpha(t)\propto \sqrt{t},
\ee
the local slope of the trajectory decreases as $J$ increases. 
This effect is, however, suppressed by $\frac{1}{N^2}$ in the 
large $N$ limit.

\subsection{High-energy scattering prediction}
\label{subsection_highE}

The contribution of the leading glueball trajectory to the total cross-section
behaves as 
\be
\Delta \sigma\propto s^{\alpha_0-\frac{1}{2}},
\ee
which means, given our calculated value $\alpha_0\simeq -1$ , that it is
suppressed as $\sim s^{-\frac{3}{2}}$. Thus the high energy
scattering of glueballs is not dominated by Regge pole exchange in 2+1
dimensions; at least if we believe that the cross section should
be constant at high energies (up to powers of $\log{s}$). 

Going back to 
section~\ref{subsection_regge}, we note that the other terms contributing to
the scattering amplitude are the isolated s-wave amplitude $a_0(s)$ and the 
background integral. Because there is a unitarity bound on each partial wave
separately, namely
\be
{\rm Im}~a_n\geq |a_n|^2,
\ee
the contribution of any partial wave amplitude to the total cross-section
is bounded by $\sim s^{-\frac{1}{2}}$. Thus this s-wave amplitude will
not dominate either at high energies. That, then, only leaves the 
background integral. If the partial wave amplitude $a(\lam,t)$ were
meromorphic in the region $0<{\rm Re}~\lam < \frac{1}{2}$, 
we would simply get additional Regge pole contributions, which should show
up as physical states by analytic continuation. Therefore there must be 
a more complicated  singularity structure in that region. For instance 
it is well known that $\lam=0$ is a logarithmic 
branch point of the partial wave amplitude
$a(\lam,t)$ at low energies (
\cite{chadan98}; 
see also appendix B). 
Also, Li and Tan
\cite{tan}
remark in their second publication that the dipole-dipole 
forward scattering amplitude
can be written as a contour integral in the complex $\lam$ plane around 
$\lam=0$:
\be
A(d,d',s)=\frac{2\pi g^2 d~ d'}{ N_c}\frac{1}{\log{s}}=
-\frac{2\pi g^2 d~ d'}{ N_c}\int\frac{d\lam}{2\pi i}s^\lam \log{\lam}
\ee
where $d,~d'$ are the sizes of the scattering dipoles; again, the 
logarithmic branch point seems to dominate the scattering process.
This intriguing similarity suggests a universal contribution from the point
$\lambda=0$.


\section{Conclusion}
\label{section_conc}

In this paper we have carried out a lattice calculation of part
of the higher spin mass spectrum of $SU(2)$ gluodynamics 
in 2+1 dimensions. Such a calculation can tell us what 
the leading glueball Regge trajectory looks like and, in
particular, whether it resembles the Pomeron. 

To provide some motivation for this question, we showed 
how simple glueball models predict
linear Regge trajectories, with small slopes, in both 2+1 and 3+1
dimensions. We emphasised that, unlike the case of $q\bar{q}$ mesons,
there are two natural models: the open adjoint string that
is the natural extension of the usual Regge model for mesons,
and the closed flux tube which has no analogue for the usual mesons,
but which arises naturally in string theory approaches to
$SU(N)$ gauge theories.
One may speculate that both models contribute and that there
are two Pomerons (for which there is some experimental evidence).

Another part of our motivation for a study in D=2+1 is an intuition 
that in high energy scattering the colliding 
glueballs should behave like `black 
segments' (analogous to the `black discs' of D=3+1) so
that the cross-section is approximately constant at high $s$.
Of course in D=2+1 we have no experimental support for such an
intuition and we therefore investigated how various field theoretic 
approaches to high energy scattering can be translated from
D=3+1 to D=2+1. The generic change is that infrared divergences
become much more severe so that, for example, one can no
longer predict directly from the BFKL equation 
\cite{BFKL2+1}
a power-like dependence
of the cross section in $s$. However there exist alternative 
analyses
\cite{tan}
done in the framework of leading-logarithmic perturbative expansion
 that do indeed obtain cross-sections 
which are constant (up to logarithms).

The framework for Regge poles is Regge theory and we saw
that there are significant changes when we go from 3 to 2 
spatial dimensions. In particular the $l=0$ partial wave
is not included in the Sommerfeld-Watson transform and
the background integral is only down by $1/\sqrt{s}$.
(In D=2+1 the intercept of a trajectory $J=\alpha(t)$ 
that gives a constant cross section is at $\alpha(0)=1/2$
in contrast to $\alpha(0)=1$ in D=3+1.) One can imagine
that the complicated singularity structure at $Re\lambda=0$,
which is not associated with particles of the theory, might
be promoted to a dominant contribution to the high
energy  cross section.

With the above background in mind, we presented the results
of our lattice calculation of the higher spin glueball spectrum.
This is a pioneering calculation and like all such
calculations can be improved upon in many respects. However we are
confident in the robustness of the results that we obtain. In 
particular, extrapolating our masses to the continuum limit shows that 
the leading Regge trajectory in the (mass)$^2$ versus spin plane
is in fact linear (to a good approximation).  Moreover it
has a small slope that lies roughly midway between the predictions
of the flux tube and adjoint string models. The intercept at $t=m^2=0$ is
$\alpha_0\simeq -1$. We identified a parallel daughter trajectory, lying about
one unit of $J$ lower.  We were also able to determine the
leading Regge trajectory for other $SU(N)$ groups and
found that the result depends very little on $N$. In 
particular it is essentially the same for the theoretically
interesting $SU(\infty)$ limit.

The very low intercept of the leading glueball trajectory
($\alpha_0\simeq -1$) indicates that the moving Regge pole corresponding 
to these glueball states gives a negligible contribution to high-energy 
scattering in 2+1 dimensions. We concluded that there must be a 
more complicated singularity
structure of the partial wave amplitude in the complex angular momentum plane
$\lam$. Evidence for a possibly universal branch point at $\lam=0$ comes
mainly from low-energy potential scattering 
(where the result is independent of the potential
\cite{chadan98}) 
and is suggested by the $\frac{1}{\log{s}}$ scattering amplitude 
found by Li and Tan in $QCD_2$ high-energy scattering.

These statements are all quite different to what one expects in 3+1
dimensions. There the phenomenological Pomeron is widely thought to be 
related to the glueball spectrum of the $SU(3)$ gauge theory, in which 
case the leading glueball trajectory had better have an intercept $\alpha_0$
around 1. We are currently performing a similar calculation in D=3+1 $SU(3)$
gluodynamics. Preliminary results
\cite{CERNtalk} 
indicate that a straight line passing  through the $2^{++}$ and
$4^{++}$ states in a Chew-Frautschi plot does not pass through 
the $0^{++}$ (in contrast to what we found
in $D=2+1$) and has parameters $\alpha_0=1.03(40)$ and 
$2\pi\sigma \alpha'=0.27(9)$; the latter translating to 
$\alpha'\simeq 0.22(8) {\mathrm GeV}^2$ if we use 
$\sigma \simeq  (0.44 {\mathrm GeV})^2$. 
These characteristics are broadly compatible with the 
well known properties of the soft Pomeron.

\section*{Acknowledgements} 
One of us (HM) thanks the Berrow Trust for financial support.
The numerical calculations were performed on PPARC and EPSRC
funded workstations in Oxford Theoretical Physics.\\
\vspace{0.5cm}
\begin{center}
 --- ${***}$ ---\\
\end{center}
\vspace{0.5cm}
\appendix

\section{Regge theory in 2+1 dimensions}

The theory of the $S$-matrix can be developed in an entirely analogous way 
to the 3+1 dimensional case, using the usual fundamental 
postulates
\cite{book}:
\begin{itemize}
\item the $S$-matrix is Lorentz invariant;
\item the $S$-matrix is unitary;
\item the $S$-matrix is an analytic function of Lorentz invariants, with
only those singularities required by unitarity.
\end{itemize}
We denote by $A$ the $a+b\rightarrow c+d$ amplitude. In 3+1 dimensions, this 
is a dimensionless quantity, whereas in 2+1 dimensions, it has unit of mass.
Therefore we define 
\be
T(s,t)=\frac{1}{\sqrt{s}}A(s,t)
\ee
The partial wave expansion in the $s$-channel reads
\be
T(s,t)= a_0(t)+2\sum_{\lam\ge1}a_\lam(t)C_\lam(1+\frac{2s}{t}).
\ee
Here 
\be 
1+\frac{2s}{t}=\cos{\theta}\qquad (s-{\rm channel})
\ee
 in the $s$ channel and 
\be
C_\lam(\cos{\theta})=\cos{\lam\theta}
\ee
 is a Chebyshev polynomial.
 The absence of a factor 2 in the first term
originates from the geometric difference between the spin 0 and the other
partial waves. If we define a parity axis along the axis of the collision, 
then  while the left- and right-winding spin $\lam$ wave functions 
add up to $2\cos{\lam\theta}$, in the spin 0 sector the negative parity state
does not have a wave function (as is the case in 3+1 dimensions); therefore
only the $0^+$ state contributes as a partial wave. This separation of the 
spin 0 sector is necessary in order to carry out the analytic continuation 
in $\lam$ through the Sommerfeld-Watson transform. 
In 2+1 dimensions, the complications due to the signature $\eta=\pm1$
also appears   since the wave functions of spin $\lam$ are associated with a 
phase $(-1)^\lam$ under a rotation by $\pi$. Thus we have to introduce two
analytic functions $a^+(\lam,t)$ and $a^-(\lam,t)$, so that
\be
T(s,t)=a_0(t)+i\int_C d\lam~\sum_\eta \frac{\eta+e^{-i\pi \lam}}{2} 
\frac{a^{(\eta)}(\lam,t)}{\sin{\pi \lam}}~C(\lam,1+\frac{2s}{t})
\ee
We now want to deform the contour as is done in 3+1 dimensions.
However because 
\be
C_\lam(z)\sim z^{|\lam|}\qquad (|z|\rightarrow\infty),
\ee
we cannot reduce the 'background integral' by pushing it to
 $\re \lam=-\frac{1}{2}$. Therefore we integrate along the imaginary axis
and  arrive at the  following expression:
\ba
T(s,t)&=&a_0(t)
+i\int_{\epsilon-i\infty}^{\epsilon+i\infty}d\lam~
\sum_\eta \frac{\eta+e^{-i\pi \lam}}{2}
\frac{a^{(\eta)}(\lam,t)}{\sin{\pi\lam}}~
C(\lam,1+\frac{2s}{t})+\nonumber\\
&&\sum_{\eta}\sum_{n}\frac{\eta+e^{-i\pi \alpha_{n_\eta}}}{2} \frac{2\pi
\mathrm{Res}_{n_\eta}(t)}{\sin{\pi\alpha_{n_\eta}(t)}}~C(\alpha_{n_\eta}(t),1+\frac{2s}{t})
\ea
So unless  $a_0(t)$  and the background integral vanish, we obtain 
\be
T(s,t)\sim s^{\mathrm{max}({\bar\alpha}(t),0)}
\ee
where  ${\bar\alpha}(t)$ is the pole with the largest real part (``leading
Regge pole''). Using the optical theorem at high energies
 \be
\sigma_{tot}=\frac{1}{s}~\mathrm{Im}~A(s,t=0)=\frac{1}{\sqrt{s}}~\mathrm{Im}~
T(s,t=0),
\ee 
we obtain the prediction, for scattering driven by  Regge-pole-exchange,
\be
\sigma_{tot}\sim s^{\bar\alpha(0)-\frac{1}{2}}
\ee
If all Regge trajectories have negative intercept, 
the background term prevails at high-energy.
In the case of potential scattering, 
$\lam=0$ is an accumulation point of 
Regge poles when $t\rightarrow 0$
\cite{chadan98}.
It is for that reason that we kept the 
background integral along $\re \lam=\epsilon$.
\newpage
\section{Potential scattering $\&$ bound states in the plane}

The Ansatz 
\be
\psi(r,\varphi)=\sum_{\lam=-\infty}^\infty \frac{\phi_\lam(r)}{\sqrt{r}}~ e^{i\lam\varphi}
\ee
plugged into the Schroedinger equation
leads to the following radial equation for $\phi_\lam$:
\be
-\phi_\lam(r)''+\left(\frac{(\lam^2-\frac{1}{4})}{r^2}+V(r)\right)\phi_\lam(r)=E\phi_\lam(r)
\ee
Thus there is a trivial correspondence between scattering in 3 dimensions and 2 dimensions
via the substitution:
\be
\ell=\lambda-\frac{1}{2}\qquad \Rightarrow \qquad \ell(\ell+1)=\lam^2-\frac{1}{4}
\ee
This effective shift in the angular momentum has important consequences. Regge originally
showed for a large class of potentials 
in 3d scattering that the partial wave amplitudes are meromorphic in $\ell$ in the 
region $\re \ell > -\frac{1}{2}$ ; this corresponds to the region $\re \lam > 0$ in 2d.
It was already known in the sixties that
at threshold $E\rightarrow 0$,
there is an accumulation of an infinite number of Regge poles around $\lam=0$.

\paragraph{The point $\lam=0$\\}

We momentarily restore the ordinary units of quantum mechanics. Because of 
the Heisenberg uncertainty principle $p_r\geq \hbar/2r$, we have
\be
E\geq \frac{\hbar^2}{8mr^2}+\frac{\hbar^2}{2m}\frac{\lam^2-\frac{1}{4}}{r^2}+V,
\ee
which at $\lam=0$ simply reduces to $E\geq V(r)$. 
Thus this exact cancellation between zero-point quantum fluctuations
implies that any attractive potential, however weak, will create a bound
state at $\lam=0$. Indeed, a heuristic calculation can be found 
in~\cite{landau} showing that the binding energy is a non-perturbative 
expression in the potential:
\be
E\simeq \exp{-\frac{1}{\int V(r) rdr}}
\ee

\paragraph{Low-energy potential scattering}

It was shown in~\cite{chadan98} that under 
very general conditions, the $s$-wave amplitude vanishes logarithmically 
at threshold. This can be interpreted as a branch point 
singularity in the complex $\lam$ 
plane:
\be
a_0\sim \frac{\pi}{2\log{k}}=\frac{\pi}{2} \int\frac{d\lam}{2\pi i} k^\lam 
\log{\lam}\qquad (k\rightarrow 0)
\ee

\newpage

\newpage
\appendix
\begin{table}
\begin{center}
\begin{tabular}{|c|c|c|c|c|c|c|c|c|}
\hline
IR& state  & $\beta=6$ &$\beta=7.2$ &$\beta=7.2$ &  $\beta=9$ & $\beta=12$ & $\beta=12$ &  $\beta=18$  \\
  &      & $ L=16 $  &$ L=20 $  & $ L=40 $  & $ L=24 $ &  $ L=32 $ & $ L=42 $&$ L=50 $ \\
\hline
 & $\sqrt{\sigma}$&$0.2538(10)^*$&0.2044(5)&  0.2072(46)    &$0.1616(6)^*$&$0.1179(5)^*$ 
 &$0.1179(5)^*$   & 0.0853(14)\\
\hline
$A_1$& $0^+$ &  1.198(25) &0.981(14)& 0.951(14)  &0.7652(78)  &  0.570(11)& 
0.577(13)&   0.3970(78) \\
&$0^{+*}$&  1.665(43) &1.396(21) & 1.394(18) &1.108(23) &0.847(18)
& 0.839(40)   & 0.584(32)   \\
&$0^{+**}$& 2.198(76) &1.859(25)&1.778(34)  &1.426(37)   &0.980(28)&  
1.00(60)   & 0.717(76)   \\
&$0^{+***}$&2.27(10)  &2.084(41) & 2.067(54) &1.522(36)   &1.226(17) & 
 1.16(12)   & 0.845(37)   \\
&$4^+$ &    2.44(27)  &2.07(33)& 2.146(64)   &1.570(39)   &1.195(47) & 
1.259(98)    & 0.798(32)   \\
&$4^{+*}$ &          &2.53(13) &   2.50(14) &1.700(52)   &   1.419(90)&  
1.500(48)   & 0.963(45)   \\  
\hline
$A_2$& $4^-$ &    2.54(12)  & 2.210(53) &2.270(64) &    & &    &    \\
\hline
$A_3$&$2^+$ & 1.957(48) &1.584(18)&1.567(18)  & 1.232(38)   &0.933(11)& 
 1.035(16) & 0.634(18)   \\
&$2^{+*}$&  2.08(18)  &1.870(37)  & 1.891(39)   & 1.421(44)   & /   & 
1.090(19) &  0.667(20)\\ 
&$2^{+**}$& 2.34(25)  &2.219(90)  &  2.242(77) & 1.660(54)   &1.152(42)  & 
1.096(92) &  0.862(14)   \\
&$2^{+***}$&2.65(29)  &2.451(71) &  2.47(12) & 1.746(56)   &1.459(29)  & 
1.385(36) &   1.019(92)   \\
&$6^{+}$&   2.93(21)  &2.51(19) & 2.64(15) & 1.878(86)   &1.438(28)  &
1.544(60) &   0.906(69)   \\
\hline
$A_4$& $2^{-}$ &    2.071(48) &  & & 1.274(37)   &0.931(24)   &  & 0.643(19)\\
&$2^{-*}$ &    2.084(56)&  & & 1.359(46)   & /   &  &  0.676(27) \\
&$2^{-**}$ &           &  & & 1.629(59) &1.222(19)    &  &   \\
&$2^{-***}$ &          &  & & 1.741(59)&    &   &  \\
&$6^{-}$ &             &  & & 1.949(64)   &1.411(21) &    &  0.952(61)  \\
\hline
\end{tabular}
\end{center}
\caption{The lightest 2+1 $SU(2)$ glueball masses on $L^3$ lattices at the
values of $\beta$ indicated. An asterisk on the string 
tension indicates that the value is taken from~\cite{teper98}.}
\la{massdata}
\end{table}
\clearpage

 

\begin{table}
\begin{center}
\begin{tabular}{|c|c|c|c|c|}
\hline
state    & $\beta=7.2$    & $\beta=9$ & $\beta=12$ &  $\beta=18$  \\
         &  $ L=20 $      &  $ L=24 $ &  $ L=32 $  &      $ L=50 $ \\
\hline
$0^+$     &  1         &  1 & 1        & 1      \\  
$0^{+*}$  &  1         &  1 & 1        & 1      \\
$0^{+**}$ &  1         &  1 & 1        & /      \\
$0^{+***}$&  0.59(12)  & 0.68(62)  & 0.97(13) &  1       \\
$4^+$     &  0.94(9)   & 0.98(4) & 0.38(2)&   0.95(2)     \\
$4^{+*}$  &    /        & 0.67(16) & 0.59(4)&   0.98(3)  \\
\hline
$2^+$ &      1   &  1         &        1 &  \\ 
$2^{+*}$&    1   &  1         &        1 &  \\ 
$2^{+**}$&   1   &  1         & 0.97(11) &  \\ 
$2^{+***}$&  1   &  1         &        1 &  \\ 
$6^{+}$&     /   &  0.87(8)  & 0.88(4)&  \\
\hline
\end{tabular}
\end{center}
\caption{The Fourier coefficients of the spin $J$ states given in table
\ref{massdata}: $|c_J|^2$  at $\beta=~7.2,~9,~12$ and 18. 
When the coefficient is larger than 0.99, we round it off to 1.}
\la{coefdata}
\end{table}

\begin{table}
\begin{center}
\begin{tabular}{|c|c|c|c|}
\hline
state & $m/\sqrt{\sigma}$ &  conf. lev. & nb. of $\beta$   \\
\hline
$0^+$ &    4.80(10)     & 78  & 5    \\
$0^{+*}$&  7.22(24)      & 64  & 4    \\
$0^{+**}$& 8.47(30)      & 17  & 5    \\
$0^{+***}$& 11.15(45)    & 14  & 3    \\
$4^+$ &    9.75(45)      & 71  & 5    \\
$4^{+*}$ &    12.06(88)      & 29  & 3    \\
\hline
$2^+$ &    7.85(15)        & 50  &5   \\
$2^{+*}$&  7.90(25)        & 15  &5   \\
$2^{+**}$& 10.00(33)       & 32  &5   \\
$2^{+***}$&13.90(71)       & 35  &3   \\
$6^{+}$&  12.09(40)        & 38  &5   \\
\hline
\end{tabular}
\end{center}
\caption{The lightest 2+1 $SU(2)$ glueball masses in the continuum limit.}
\la{contdata}
\end{table}
\begin{figure}[bt]

\centerline{\begin{minipage}[c]{13cm}
    \psfig{file=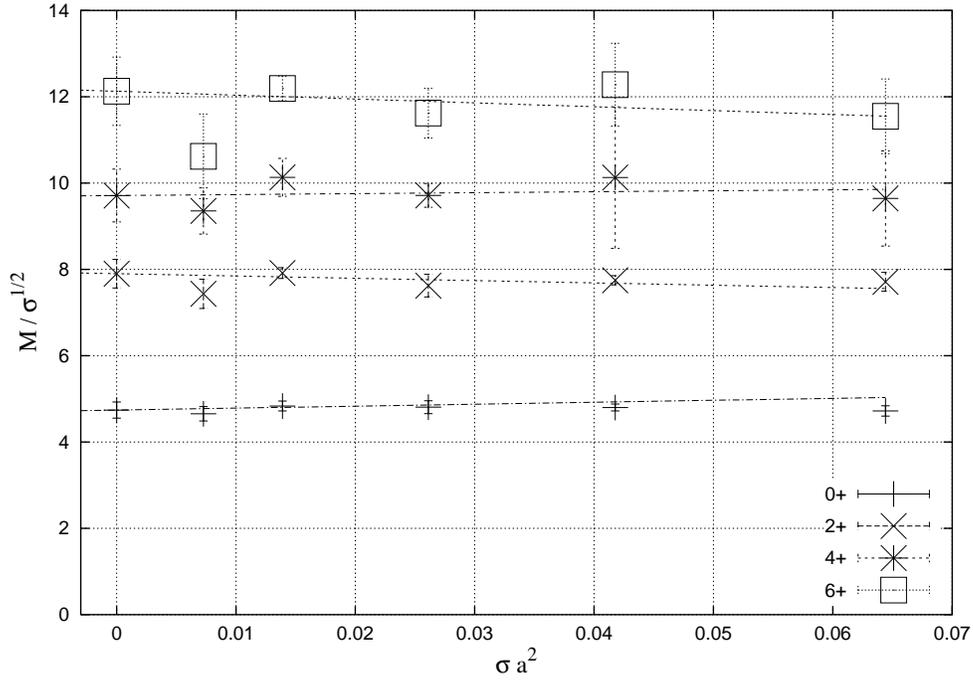,angle=0,width=13cm}
    \end{minipage}}
\vspace*{0.5cm}
\caption[a]{The continuum extrapolation of the lightest 2+1 $SU(2)$ glueball
 in the $0^+,~2^+,~4^+$ and $6^+$ sectors. The points at $a^2\sigma=0$ 
represent the result of the continuum extrapolation.}
\la{extrapolfig}
\end{figure}
\begin{figure}[bt]

\centerline{\begin{minipage}[c]{13cm}
    \psfig{file=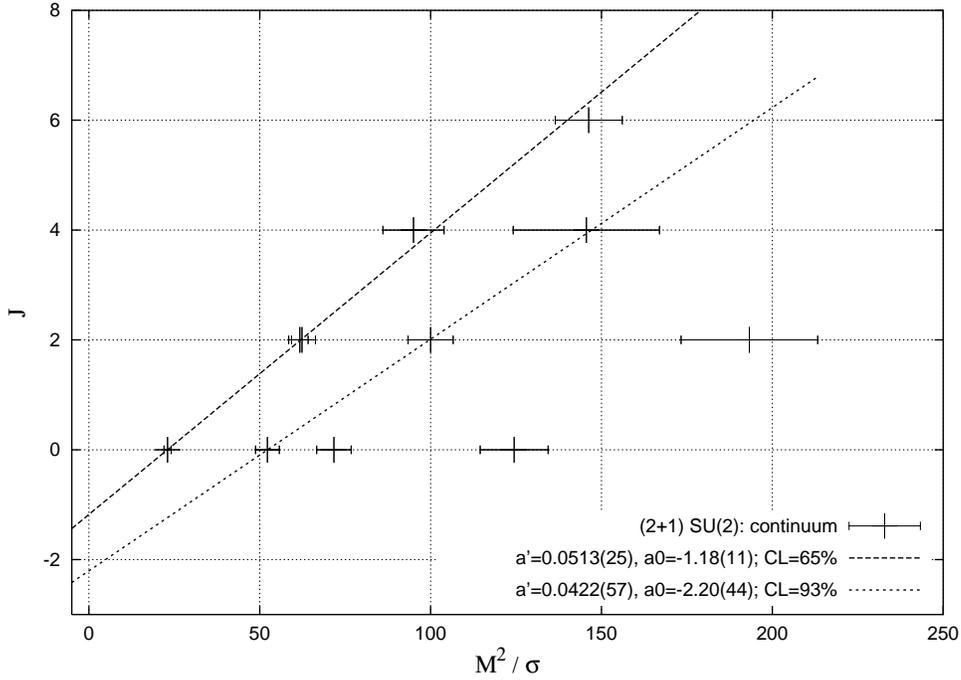,angle=0,width=13cm}
    \end{minipage}}
\vspace*{0.5cm}
\caption[a]{The Chew-Frautschi plot of the continuum  2+1 $SU(2)$ 
glueball spectrum}
\la{chewfig}
\end{figure}
\begin{figure}[bt]

\centerline{\begin{minipage}[c]{13cm}
    \psfig{file=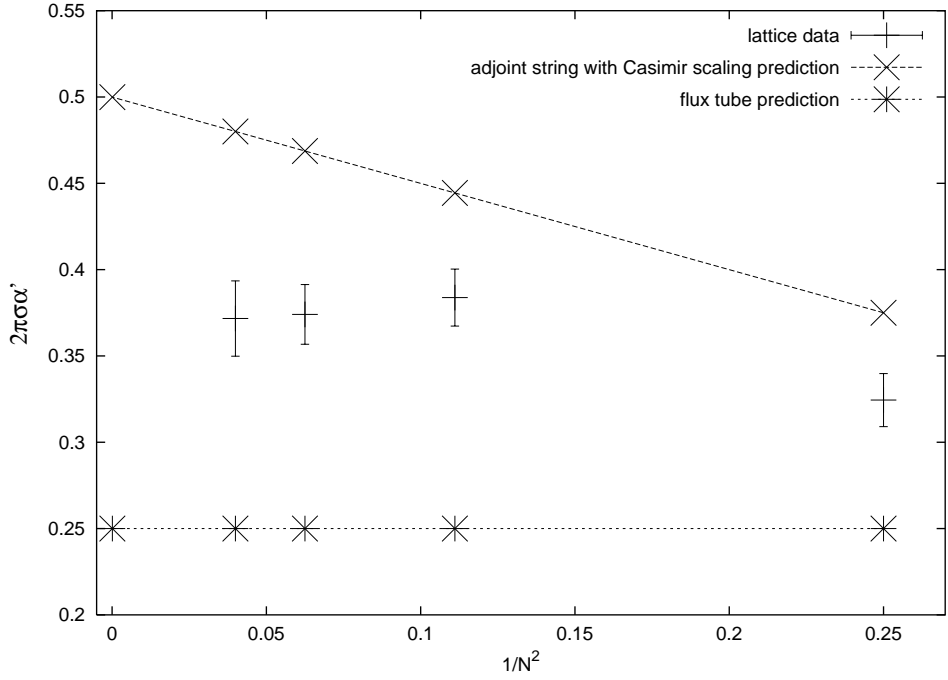,angle=0,width=13cm}
    \end{minipage}}
\vspace*{0.5cm}
\caption[a]{The slope $\alpha'$ of the leading Regge trajectory in 2+1 $SU(N)$
gauge theory, in units of
$\frac{1}{2\pi\sigma}$, as a function of $\frac{1}{N^2}$}
\la{a1}
\end{figure}
\begin{figure}[bt]

\centerline{\begin{minipage}[c]{13cm}
    \psfig{file=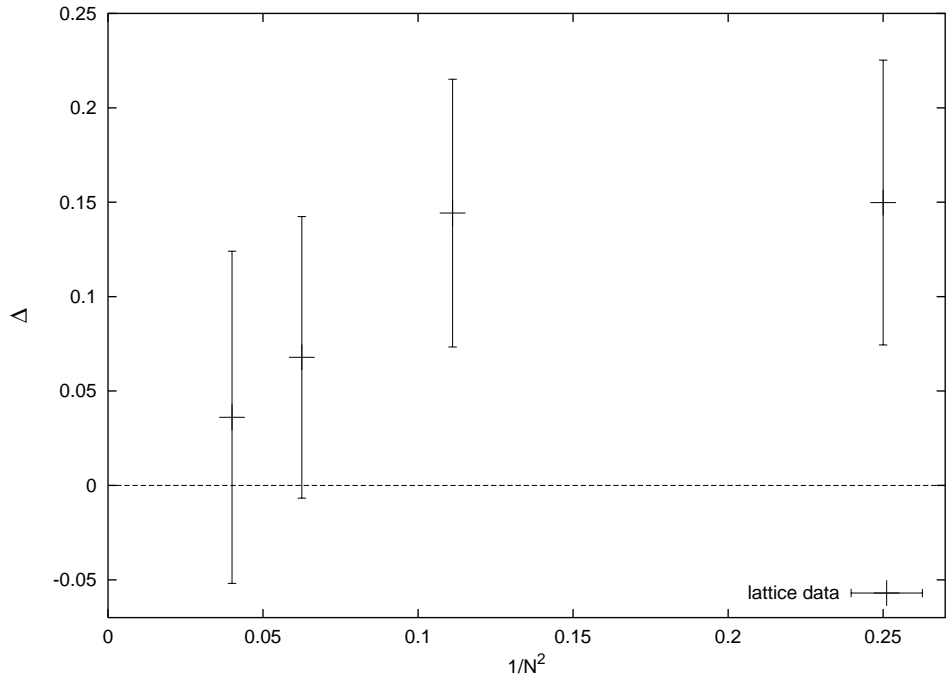,angle=0,width=13cm}
    \end{minipage}}
\vspace*{0.5cm}
\caption[a]{The difference $\Delta\equiv -(\alpha_0+1)$ of the intercept to 
the value -1, as a function of $\frac{1}{N^2}$ in 2+1 $SU(N)$
gauge theory}
\la{a0}
\end{figure}
\end{document}